# A system of different layers of abstraction for artificial intelligence

*Alexander Serb, Themistoklis Prodromakis*

*Univeristy of Southampton, UK*

**The field of artificial intelligence (AI) represents an enormous endeavour of humankind that is currently transforming our societies down to their very foundations. Its task, building truly intelligent systems, is underpinned by a vast array of subfields ranging from the development of new electronic components to mathematical formulations of highly abstract and complex reasoning. This breadth of subfields renders it often difficult to understand how they all fit together into a bigger picture and hides the multi-faceted, multi-layered conceptual structure that in a sense can be said to be what AI truly is. In this perspective we propose a system of five levels/layers of abstraction that underpin many AI implementations. We further posit that each layer is subject to a complexity-performance trade-off whilst different layers are interlocked with one another in a control-complexity trade-off. This overview provides a conceptual map that can help to identify how and where innovation should be targeted in order to achieve different levels of functionality, assure them for safety, optimise performance under various operating constraints and map the opportunity space for social and economic exploitation.**

AI is currently transforming our society, affecting every aspect of our lives. From automated finance and intelligent security to image recognition and natural language processing machine intelligence is disrupting the way we do business and carry out our daily lives. However, despite the unquestionable advances in the field, the main understanding of what constitutes AI tends to revolve around a simple combination: the algorithms of statistical learning and the implementation of multi-layered artificial neural networks (ANNs). Simultaneously, hardware developers generally focus on implementing the connectomics of ANNs and strive to accelerate their operation. This fragmented and compartmentalised view of AI, however, does not exploit the full potential of the field. We would argue that a holistic understanding of AI requires and awareness of how intelligent systems can be interpreted at different levels of abstraction. Future innovation is set to greatly benefit from an ability to facilely switch between interpretations. In this work we intend to: First, provide an overview of the different fields that collectively constitute the construct of AI. Second, show how these fit into a hierarchical structure. Third, explain the interaction between different LoAs. Fourth, discuss the internal structure of each level.

## Levels of abstraction

Any system can be regarded as a complex structure which contains one or more LoAs. We may define a LoA as a: 'functionally complete ontology that describes the workings of a system under study'. More elaborately: a collection of terms, statements and guesses used to characterise the behaviour and function of a system, explaining as much as is relevant in view of the terms being used. To give a concrete example, the function of a car can be explained in terms of the functionality of the steering wheel and pedals; an explanation very useful to the prospective driver. However, an alternative explanation can be provided in terms of the functionality of individual mechanical components, such as would be useful for a car mechanic (and less so for the driver). Constructing ontologies describing systems at higher LoAs is arguably absolutely essential for managing the complexity of large systems using human-level intellectual resources. The price, however, is that each LoA hides complexity and detail present in the layer below; a reward/effort trade-off. From this we can infer that artificially recreating a highly complex phenomenon such as intelligence drives us to push the boundaries of our comprehension and engineering capabilities to ever higher LoAs. This has already been demonstrated successfully by the humble personal computer (PC): microprocessors of today routinely contain billions of transistors that a human engineer can only comprehend and handle when bundled in large modules.

Nevertheless, the occlusion of complexity is not the only interaction between levels. Equally importantly, a well-designed (or evolved) LoA will fortify itself against uncertainty factors present at lower levels. Thus, it is not only detail that is lost when climbing up the ladder of the hierarchy, but also uncertainty. Perhaps one of the best examples is found when moving up from single-transistor thinking to logic gate thinking. Arguably, one of the key factors that has led to the near-complete dominance of digital over analogue is the fact that in digital domain transistors are operated in such manner that they successfully approximate simple switches. A full analogue model of a transistor (e.g. the models of the BSIM family) may have 300+ parameters [1], but once a logic gate is committed to layout the amount of useful information for design is reduced to a truth table and a standard delay file describing the latencies from inputs to outputs. This methodology propagates further up the chain, with typical examples including glitch-proofing of logic gates through judicious system design [2], [3] and ECC (error-correction coding) for communication channels [4].

LoA partitionings may take many forms, but in practice some are more natural and useful than others. As practical guideline good partitionings will: First, minimise connections and interactions (e.g. signals) between levels. Second, maintain alignment of the borders of higher level LoAs with lower level LoAs. As an example a module-level description of a processor should ideally not 'cut' its 32-bit ALU (a lower level module functioning as a cohesive 32-bit unit) into a 7-bit lower part and a 25-bit upper part (unless there is a very specific reason that the borders between LoAs do not align). These guidelines are very natural for traditional computing systems, but the picture is nowhere near as clear in neuro-inspired AI systems. When we conduct our electrophysiology experiments or formulate our theoretical hypotheses for the function of brain circuits how do we know that we are not trying to explain functionality 'in units of 1.5 gates'?

## The hierarchical structure of AI systems

In this work we use a five-LoA partitioning of AI systems starting from the agency layer at the high abstraction end, and finishing with the physical (implementation) layer at the low end. Our breakdown carries echoes of Marr's tri-level hypothesis [5], where a conceptual split between computation (motivation and problem posing), algorithmics (representation and processes) and implementation (structures and signalling) was provided. However, whilst Marr's partitioning is more abstract and flexible, (is object recognition a computational problem, or a process to serve a higher purpose?) we propose a much more specific and rigid framework which we consider particularly useful as a thinking tool for guiding innovation (whilst keeping firmly in mind that it is only one of many possible options). Within this framework we link each layer to an ontology and determine the nature of its signals, computational units and relationship to learning. The five-LoA structure is specified as follows:

Floor 5 - The agency layer: This level is concerned with issues such as: how AI systems reason and formulate evaluations of their current situations, how goals are formulated and how behaviour is elicited towards reaching said goals. Systems are described in terms of complex semantic meanings, goals and emotions, which is relevant to researchers in fields such as psychology, education and multi-agent systems [6]. A number of cognitive architectures/systems have been developed over the years [7] with notable examples found in the SPA [8] and ACT-R [9].

At such high level the process of learning is largely behavioural and educational; learning processes target desired behaviours and equip the system with the high-level declarative knowledge required to successfully execute them. These objectives relate to the fields of reinforcement learning and knowledge representation [10] and for sufficiently intelligent systems may start spilling over into fields such as educational psychology.

Finally, we note that this layer generates behaviour as an output. This is of direct interest to areas such as human-machine and machine-machine interactions/interface design, but also to assurance and ethics (particularly in light of recent research demonstrating that socially acceptable ethical norms may vary substantially between cultures [11], especially when ambiguous information is involved).

Floor 4 – The semantic layer: Here the system manipulates symbols, i.e. representations of the outside world, internal states and memories in order to perform reasoning and/or planning tasks. At this layer systems are discussed in terms of high-level operations such as variable binding [12], inference, memory storage, memory recall etc. The high-level 'cognitive modules' executing these operations communicate using the language of semantic objects, a.k.a. 'symbols', which are typically encoded as vectors (see [13] for an example of vector representation of words). Vector encoding enables semantic embedding, i.e. capturing the 'relatedness' of semantic objects via a distance metric [14].

Learning at the level of semantic object manipulation includes such important effects as 'chunking' [15], whereby frequently used or otherwise salient combinations of semantic objects are turned into new semantic objects. This is proposed as a mechanism for constructing ever more abstract concepts in the mind and ultimately giving rise to intelligence [16]. Whist chunking can be regarded as a straightforward extension of the ever higher level features created by deep neural networks (DNNs), AI systems with powerful reasoning capabilities possess the tools to combine symbols in novel ways and thus create truly new semantic objects (e.g. 'blue moon' or 'griffon'). Notably, such complex combinations of semantic objects are underpinned by different mathematical machinery as compared to the standard feature abstraction performed by DNNs (see [17] for a comprehensive treatise on semantic-layer processing).

Floor 3 – The computational layer: At this stage more basic functions are carried out at below semantic object level. These might include classification/identification of sensory inputs [18], gain control of auditory signals/contrast enhancement of visual inputs [19], [20], control loops/signal generation or modulation for supervised learning (e.g. olivo-cerebellar system [21]) and more. Such modules communicate with each other using abstract vectors/numbers, which may be encoded as $n$-bit digital numbers [22], neural spikes (of different levels of complexity - see [23] for an example of different 'flavours' of spikes) or in real-valued analogue [24]. Ultimately they handle vector information, where each vector may or may not constitute a semantic object. At this level systems are described in terms of neurons and layers, partially or fully connected and engineering problems pertain to classification accuracy, local circuitry stability, and fast efficient training. These are addressed using a multitude of tools such as: ANN architecture development (ANN layer number, size and connectivity; use of feedforward [25]–[27] and feedback [28]–[30] links), efficient filter design (e.g. convolutional neural network [27] kernels for image processing), resource allocation towards representing abstract numbers (bit-width, number of neurons used to represent an abstract number [17] etc.) and careful choreographing of various learning rules [31].

The terminology used to describe learning at this level is already widely familiar: supervised and unsupervised learning and the associated algorithms used to implement them, with the error backpropagation [32] and 'winner-take-all' techniques [33] very widespread. The bibliography on learning is extremely extensive, but virtually all –if not all- techniques essentially perform gradient descent optimisation.

Floor 2 – the functional layer: This consists of the fundamental mathematical function implementation blocks. Systems are therefore analysed in terms of logic (or threshold logic) gates, shift registers, analogue RC circuits etc., or even individual transistors; all often used to implement artificial neurons or parts thereof. At this level signalling occurs at below the abstract numerical object level; in terms of voltage/current signals and waveforms. It is also at this level that issues such as artificial neuron activation function shape [34]–[36], somatic behaviour (e.g. spiking threshold, intrinsic neuronal plasticity) [37]–[39] and synaptic behaviour (input/output transfer characteristic, temporal dynamics, learning rule etc.) [40]–[43] are tackled.

Learning is a key design consideration in this layer, for it is here that the learning rules are crafted. The literature on learning rules is vast and includes rules designed for spiking ANNs (SNNs) [42], [44], [45], for non-spiking ANNs [46], which also exist in stochastic versions [47], [48], amongst other [49]. Finally, advanced versions of plasticity also exist whereby the plasticity rule itself is modified by the stream of data inputs. This is called 'metaplasticity' and effectively is the idea of equipping synapses with additional state variables beyond the synaptic weight itself [50], [51].

Floor 1 – the physical layer: At the bottom of the pyramid lies the 'physicochemical layer', or 'physical layer', whereby the language is that of moving electrons and/or ions; a world of ion channels and MOSFET channels, where the full BSIM model [1] of the transistor applies. This is the layer that most intimately relates to the physical fabrication of the system and describes systems in terms of physics, electrochemistry and fabrication processes. Floor 1 has access to the finest degree of detail, but is also exposed to most of the uncertainty; issues such as classical PVT (process/mismatch, power supply voltage and temperature) variations [52], [53] that are generally not visible beyond floor 2. Technologies such as quantum computing, optical computing, memristive devices [54] and other unconventional computational substrates are seeking to disrupt the AI hierarchy from this level up.

Learning, typically tackled at floor 2 and (e.g. synaptic weights stored digitally at the circuit level [55]) is now attacked at floor 1 by developing new artificial synapses [56]–[60] and optimising both their properties (supported weight resolution [61], scalability [62] and power efficiency [63], etc.) and their operating regimes [64]. Finally, we note that improved designs of artificial synapses using basic electronic components (i.e. sub-circuit-level solutions) have also been attempted before by using floating gate MOSFETs [65].

Finally, we note that whilst the language of the discourse might differ significantly across layers, a number of common issues reverberate across the entire hierarchy and preoccupy every level. These are issues such as handling power dissipation, uncertainty and system controllability. As an example, the question of power dissipation mainly revolves around managing system complexity at higher LoAs and unit cell efficiency at lower levels. The layered structure of AI systems is summarised in Table 1.

*Table 1: Summary of levels of abstraction in AI systems. Note that the 'internal data structures' and 'data output' columns form an escalator whereby the output of a layer forms the basis for the input to the next layer above. An easy way to read the 4th, 5th and 6th columns is: "<Internal signals> are interpreted as <internal data structures>, and many of those together give rise to <data output>", so for example: "Artificial neuron activities are interpreted as abstract numerical objects, and many of those together give rise to semantic objects". Finally, note how at the agency layer there are no subunits and no internal signals.*

| Lvl | Layer | Subunits | Internal signals | Internal data strcutures | Data output | Learning effects | Learning methods |
|---|---|---|---|---|---|---|---|
| 5 | Agency | N/A | N/A | Interpreted sentences and meanings | Behaviour | Educational, goal setting (incl. ethics) | Reinforcement |
| 4 | Semantic | Declarative memory, variable binder, working memory, semantic object encoding/grounding etc. | ANN ensemble activities | Semantic objects, either unitary or complex (e.g. sentences or sentence-level logical structures) | Interpretations of sentences / meanings | Novel combinations of abstract semantic objects, plans of action (transformations between semantic objects). | Supervised, exploration, imagination |
| 3 | Computational | ANN layers of all levels and patterns of connectivity | Artificial neuron activities | Sensory data, internal states, ANN layer outputs & other abstract numerical objects | Semantic objects | Statistical relationship capture | Supervised / unsupervised |
| 2 | Functional | Gates, artificial neurons, small, specialised circuit modules | Electrical or chemical signals | Spikes, pulses and other waveforms | Abstract numerical objects | Function reconfiguration | Application of learning rules |
| 1 | Physical | Transistors, resistors, capacitors, memristors | Electron or ion movement | Waveform segments | Waveforms | Synaptic weight (or equivalent) change | Physical changes in devices |

# What's inside a level of abstraction?

Levels of abstraction may be themselves be engineered to different levels of capability (and complexity). Let us briefly examine how this applies to each layer:

Floor 5: A putative cognitive agent could be endowed with anything between a bare minimum of functionality including limited knowledge and few rules for performing a narrow task very well (akin to expert systems, but with less explicitly descriptive rules), or it could be equipped with a learning centre allowing it to acquire more knowledge, a judgement centre allowing it to make choices in the face of ambiguity or even an 'emotion' centre providing a wholly internal driver of behaviour.

Floor 4: AI systems with more/less working memory capacity and stronger/weaker ability to combine multiple semantic objects into new entities could be envisaged. Other options might include systems with mini-Von Neumann processors embedded into their overall architecture, the ability for simultaneously handling information in parallel (multiple foci of attention with some sort of central coordinator), etc.

Floor 3: Here we already see an enormous zoo of ANN architectures including LSTMs, hierarchical memory networks, deep learners, convolutional ANNs and a myriad others. To this we might add inference engines (i.e. accelerators for computing Bayes' rule) [66], audio/visual signal pre-processors and more. Systems may be built using either more or fewer, either more or less complex ANNs belonging to these families.

Floor 2: At this level we encounter anything from the simple logic gates currently dominating the market, to artificial neurons that are either digital, or analogue, electronic synapses. In particular artificial neurons may be anything between simple integrating capacitors (or digital accumulators) to complex compartmental models featuring even intrinsic neuronal plasticity.

Floor 1: At the bottom level there has been relatively little movement; transistors have dominated the floor for decades. However, different flavours of transistors, including organics, memristive technologies ranging from simple tuneable resistances encoding synaptic weight to complex volatile/non-volatile devices capable of metaplasticity, and finally potential quantum components are promising a revolution in the field. Particularly exciting opportunities lie in high computational power synapses exploiting the intricate internal dynamics of electrochemically/magneto-resistively active devices.

The full structure of the AI system hierarchy, as used in this particular work is illustrated in Figure 1.

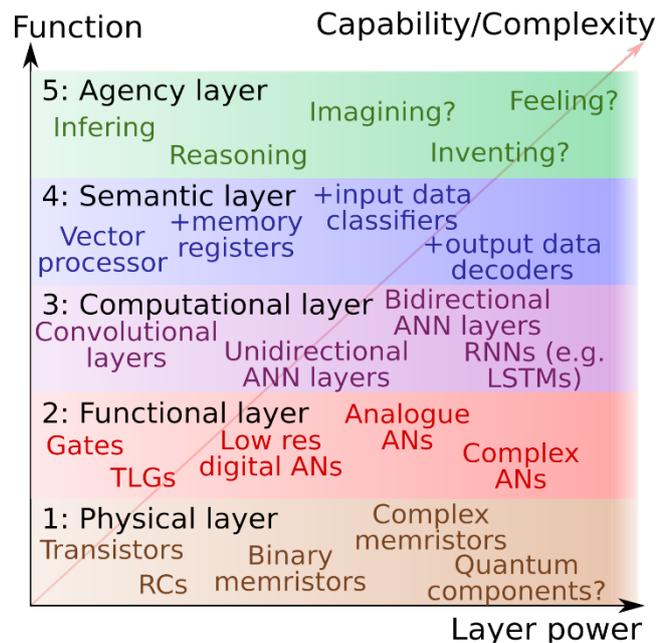

*Figure 1: Partitioning of AI systems along function axis (vertical – layers of abstraction) and along the power, i.e. the capabilities within each layer (horizontal axis). Systems towards the top right corner are more complex, but also exhibit higher processing capabilities.*

## How do LoAs interact with each other?

In general as more capability and complexity is pushed down towards lower layers we tend to obtain five effects: First, better power efficiency: This is the foundation of the hardware accelerator concept. It is more energy-efficient to implement a neuron directly on hardware rather than in software. Second: Lower complexity at higher levels. For example if complex synaptic dynamics are handled by elaborate circuitry at layer 2 the overarching ANN at layer 3 may potentially achieve the same functionality with a substantially different design (this concept was exemplified in embryonic form in [64]). Third, AI systems become more opaque. It is harder to query the internal state of a neuron (e.g. membrane capacitance voltage) when this is implemented on a physical capacitor vs. a software implementation. Fourth: reliability drops. The beneficial effects of removing uncertainty by moving up the LoA ladder become less applicable. Finally: there is less room for operational flexibility in each layer. Attempting to control system behaviour by directly manipulating semantic objects is much easier than by manipulating spike trains. The above implies that AI systems must be regarded as complex, multi-level hierarchies where each layer's complexities, capacity, benefits and flaws can be traded against other layers'.

Currently, the main focus of innovation is mostly clustered around the bottom three floors. As we have seen in previous sections, there are large communities tackling (semi-independently) the development of novel NN architectures, an array of hardware accelerators for neuro-inspired computation and novel devices for emulating synapses. This situation places accelerator designers (floor 2) in the spotlight, as they need to both cater to the needs of the NN architects (floor 3) and capture the opportunities offered by pushing complexity towards the individual device level (floor 1).

When faced with the requirements of floor 3, hardware designers have brought forth three general approaches: First, Von-Neumann bus communication as seen in e.g. Intel's Movidius [67] or NVidia's Volta system [68]. Second, Ethernet-inspired connectivity using routing tables as might be found inside Intel's Loihi [69] or the Univ. of Manchester's SpiNNaker system [70]. Third, address-event representation (AER) with hierarchical routing as found in AiCTX's DyNAP family of systems [71]. In each case the designers attempted to offer the maximum possible flexibility to their systems by allowing effectively any-to-any connectivity. Once, however, certain NN topologies become sufficiently developed and find widespread use, hardware can start being designed around the specific needs of said topologies. This is already the case for convolutional layers, with systems such as Loihi (but not only) already including shared memory blocks for the convolutional kernel weights [69]. This is a very direct example of how layers 2 and 3 may interact in a substantially impactful way.

Gazing in the opposite direction, we observe that there is a strong undercurrent of research activity in developing AI hardware using memristive synapses. This is underpinned by a strong expectation that memristive crossbars [72] will provide a solution to the persistent problem of physically instantiating vast numbers of artificial synapses within highly constrained power and area budgets. The projected benefits of using memristive synapses are going to be partially counterbalanced by the need to design and implement circuitry that can operate them using appropriate biasing schemes, because memristive devices effectively dictate the waveforms they require in order to operate as well-behaved synapses (e.g. see [73], [74]). Simultaneously, hardware designers from floor 2 put pressure on device developers to build memristors that can operate well under simple and easily implementable biasing schemes (e.g. IBM's quest for memristors that support linear resistance –i.e. synaptic weight- increments/decrements [75]). This is a very distinct example of how layers 1 and 2 interact, illustrating the significant impact that innovations in floor 1 can exert on the designs developed in floor 2 (IBM has developed a memristor-based, experimental AI chip [76] despite already possessing an advanced system based on conventional hardware in the form of the TrueNorth chip [77]).

## Concluding remarks

The field of practical AI system development is truly vast. It encompasses a staggeringly broad variety of specialisations and elements, enmeshed in a fabric so extensive that professional communication between its extremities is difficult and still limited. However, regarded from a distance this fabric can be understood in terms of a nested hierarchical structure, featuring different levels of abstraction, levels of capability/complexity within each layer and internal design hierarchies for each individual design.

In this work we have attempted to reveal this structure, or perhaps more accurately, to describe the field in terms of this structure. As a result the vastness of the field is mapped out and a simultaneous awareness of all its facets becomes possible. This is useful for revealing how the field may progress in the future, which actors might need to be brought together and collaborate to that purpose and what trends and patterns underlie the superstructure. As an example we observe that engineering efforts at lower levels of the hierarchy are particularly well-suited for meeting power optimisation requirements; efforts that could benefit from the expertise of material scientists. Conversely, higher level research is best suited to fulfilling functional capacity needs; an area that psychologists may be able to contribute to alongside scientists more traditionally linked to AI. Finally, we note that no claim is laid that this is the only valid division of the field; merely a potentially very useful one.

In conclusion, the field of physically implemented AI systems promises great opportunities and indeed excitement and progress for a long period to come.